\documentclass[twocolumn,showpacs,preprintnumbers,prl]{revtex4}

\newcommand{\BE}{\begin{equation}}
\newcommand{\EE}{\end{equation}}
\newcommand{\BA}{\begin{eqnarray}}
\newcommand{\EA}{\end{eqnarray}}
\input epsf.tex

\usepackage{graphicx}
\usepackage{dcolumn}
\usepackage{bm}

\begin{document}

\title{Sonic crystal lenses that obey Lensmaker's formula}

\author{Chao-Hsien Kuo}
\author{Zhen Ye}\email{zhen@phy.ncu.edu.tw} \affiliation{Wave Phenomena
Laboratory, Department of Physics, National Central University,
Chungli, Taiwan 32054, R. O. C.}

\date{\today}

\begin{abstract}

This paper presents a theoretical study of the phenomenon of
acoustic imaging by sonic crystals, which are made of
two-dimensional regular arrays of rigid cylinders placed in
parallel in air. The scattering of acoustic waves is computed
using the standard multiple scattering theory, and the band
structures are computed by the plane-wave expansion method. It is
shown that properly arranged arrays not only can behave as
acoustic lenses, but also the focusing effect can be well
described by Lensmaker's formula. Possible applications are also
discussed.

\end{abstract}

 \maketitle

When propagating through periodic structures, waves reveal a
particularly important property. That is, the propagation of waves
will be modulated by the periodic structures. As a result, the
dispersion of waves will no longer behave as in a free space, and
the so called band structures appear.

The phenomenon of band structures was first investigated for
electronic systems and was put on a solid foundation in the
context of Bloch's theorem\cite{Bloch}. Since the central physics
behind electronic band structures lies in the wave nature of
electrons, by analogy an immediate inquiry would naturally be
whether such a phenomenon can be demonstrated in classical systems
involving, for instance, acoustic and electromagnetic (EM) waves.
Although this problem was early addressed by Brillouin \cite{Bri},
then by Yariv and Yeh\cite{Yariv}, this seemingly straightforward
question has not received serious attention, until late 1980s. In
1987, Yablonovitch \cite{Yab} and John \cite{John} proposed that
such a band structure is indeed possible when EM waves travel in
periodic dielectric media. Since then, the study of EM waves in
periodic structures has been booming, eventually leading to the
exploration of acoustic waves in periodic structures (e.~g.
Refs.~\cite{Dowling,sanchez1,sanchez2,Kush,Ye2,Ye3}). This has
induced the establishment of the field of sonic crystals (SCs).

The exciting phenomenon of band structures in sonic crystals
allows for many possible applications. It has been recognized that
SCs could be used as sound shields and acoustic filters
\cite{sanchez1,sanchez2,caba1,rober,sanchis,kuswa}. These
applications mostly rely on the existence of complete sonic
bandgaps in which acoustic waves are prohibited from transmission
in all directions. Thus most of early studies have been focused on
the formation of bandgaps.

Recently, interest in the low frequency region, in which the
dispersion relation is more or less linear, has started. Since the
wavelength in this region is very large compared to the lattice
constant, the wave sees the media as if it were homogeneous.
Consequently, it has been proposed that refractive devices could
be developed to converge acoustic waves by SCs, that is the sonic
crystal lenses. A necessary condition to be satisfied to construct
a sonic crystal lens is that the acoustic impedance contrast
between the SC and the medium should not be large; otherwise
acoustic waves will be mostly reflected. Once this condition is
satisfied, the converging lens can be either convex or concave
depending on whether the sound speed in the SCs is smaller or
greater than that in the medium.

Recently, it was experimentally demonstrated \cite{cervera} that
proper SCs can indeed make refractive devices such as Fabry-Perot
interferometers and acoustically converging lenses. This
interesting observation has stimulated further explorations of
acoustic imaging by SCs\cite{Sanchez,Garcia,Ye4}, and has been
also extended to the optical imaging by photonic crystal
lenses\cite{Halevi,Ye5}. In light of these novel developments, in
this paper we wish to explore further the properties of the sonic
crystal made lenses.

This paper presents a first attempt to quantify the imaging effect
of sonic crystals. Here, we carry out numerical simulations on the
focusing of acoustic waves by SCs. To comply with the current
experiments\cite{cervera} so that experimental verifications could
be readily done, we will consider the SCs made of arrays of rigid
cylinders in air. It will become clear that the present approach
can be readily extended to other SCs and also to photonic
crystals. The results will demonstrate that SCs can make
refractive lenses which nicely obey Lensmaker's formula. It is
also shown that the refraction dominates the focusing properties.

Before continuing, it is worth noting here that pursuing acoustic
lenses can be dated back to 70s. For example, ultrasonic lenses
made from bulk materials were proposed by Beaver et al.
\cite{Beaver} and Szilard et al. \cite{Szi}. However, these lenses
are quite different from the sonic crystal lenses. The essential
principles that govern the focusing phenomenon by sonic crystals
are (1) the wave propagation are subdued by the crystal structure,
leading to the refraction of waves commonly seen at the interface
of two different materials; yet (2) the frequencies of travelling
waves are still within the passing band, and therefore no
propagation inhibition will incur, allowing efficient transmission
through the crystal\cite{cervera}.

Consider $N$ straight cylinders located at $\vec{r_i}$ with $i=1,
2, ...N$ to form regular arrays. An acoustic line source
transmitting monochromatic waves is placed at $\vec{r_s}$. The
scattered wave from each cylinder is a linear response to the
incident waves which are composed of the direct wave from the
source and the multiply scattered waves from other cylinders, and
subsequently contributes to the total waves, forming a
self-consistent multiple scattering scheme. The final wave
reaching a receiver located at $\vec{r}_r$ is the sum of the
direct wave from the source and the scattered waves from all the
cylinders. Such a scattering process has been formulated exactly
by Twersky \cite{twersky}, and was detailed in
Refs~\cite{Ye2,Ye3}. For regular arrays of cylinders, the band
structures are computed by the plane-wave method to obtain the
acoustic phase speed inside the arrays.

In the following computation, the rigid cylinders of radius $a$
are arranged to form triangular arrays in air with lattice
constant $d$. The following parameters from the
experiment\cite{cervera} are used in the simulations: (1) the
radius of the cylindrical rods is $2.0$cm; (2) the lattice
constant is $6.35$cm; (3) The filling factor, defined as the area
occupation of the cylinders per unit area $f_s =
(\pi/2\sqrt3)(d/a)^2$, is equal to 0.3598; (4) the sound speed in
air is $v_a = 351$m/s.  Additionally, the acoustic transmission
through the cylinder arrays is normalized as $T = p/p_0$, and the
incident wave in this entire project is along the $\Gamma X$
direction. We note here that the rigid cylinders can be any solid
rods. As shown in Ref.~\cite{sanchez2}, any material whose
acoustic impedance with respect to the air exceeds roughly 10 can
be used as the composition material for the rods.

\begin{center}
\begin{figure}[hbt]
\vspace{10pt} \epsfxsize=2.5in\epsffile{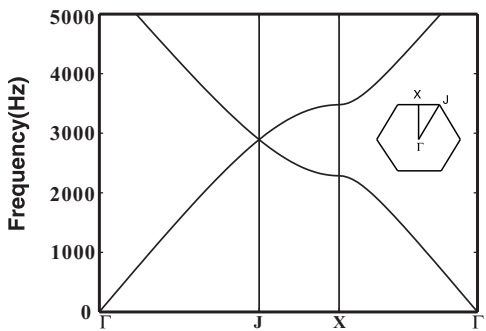} \caption{The
band structures computed by the plane wave expansion method for
triangular lattice of the rigid cylinders in air. } \label{fig1}
\end{figure}
\end{center}

Figure~\ref{fig1} presents the band structure result for the
triangular array of rigid cylinders. We see that the dispersion in
the low frequency region is linear. Our subsequent calculations
will be restricted to this region only. We calculate the sound
phase speed from the linear dispersive curve as $v_{p,c} =
\omega/K = 295$m/s. Then the refraction index is calculated as $n
= v_a/v_{p,c} = 1.1898$. We note that when waves propagate in
structure media, it is usually necessary to distinguish the group
velocity, roughly characterizing the energy flow, and the phase
velocity. In the linear dispersion regions, the two velocities are
nearly the same. We have also computed the phase velocity in other
directions, yielding the consistent values. These results imply
that the cylinder arrays may be regarded as an effective
refractive medium, making it possible that SCs can be used as the
refractive acoustic devices. This will be supported by the
following simulation of the sonic crystal lenses.

\begin{center}
\begin{figure}[hbt]
\vspace{10pt} \epsfxsize=3.25in\epsffile{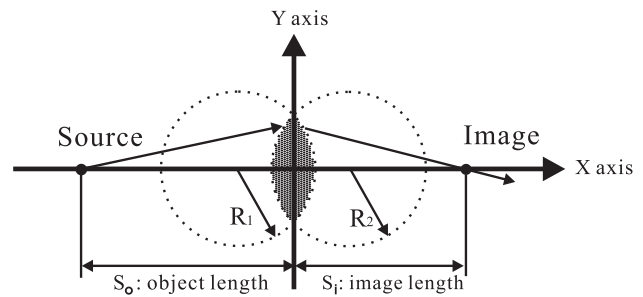} \caption{The
conceptual layout of an acoustic convex lens, in line with the
optical lens depicted in Fig.~5.16 of Hecht\cite{Hecht}}
\label{fig2}
\end{figure}
\end{center}

A conceptual layout of the acoustic lensing system made from a
sonic crystal is shown by Fig.~\ref{fig1}. The geometrical
factors, important to the following discussion, are clearly
indicated in the figure. The source is place at a distance $s_o$
from the lens. The image is at a distance of $s_i$. The cylinders,
represented by the black dots, are placed insider the lenticular
area. Inside the area, the cylinders are arranged to form the
triangular lattice with the parameters given above.
Fig.~\ref{fig2} actually shows a double two-dimensional convex
lens of a refractive sonic crystal whose surfaces have radii of
curvature $R_1$ and $R_2$. We note that a three-dimensional lens
can be fabricated by aligning two perpendicular two-dimensional
lenses.

\begin{center}
\begin{figure}[hbt]
\vspace{10pt} \epsfxsize=3in\epsffile{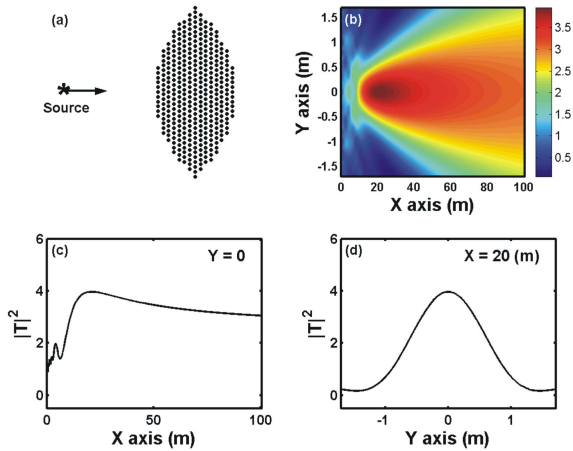} \caption{(a) The
line source denoted by {\bf S} and the lenticular arrangement of
rigid cylinders; (b) two dimensional spatial distribution of the
transmitted intensity ($|T|^2$) on the right hand side of the
crystal; (c) the variation of the intensity along the $x$ axis at
$y$=0 and (d) the variation of the intensity  along the $y$ axis
at $x$=20m. Parameters: (1) the source is place at 21m away from
the center of the crystal; (2) the origin is placed at the
rightmost point of the crystal; (3) the frequency is taken as
1500Hz, and the incident wave is along the $\Gamma X$ direction;
(4) the radii of the curvatures of the lens are chosen to be
equal, i.e. $R_1 = R_2 = 4$m. The thickness of the lens is about
80 cm or 12 lattice constants, while the height amounts to 340 cm
or 53 lattice constants; therefore it is a thin lens. The total
number of scatterers is 546.} \label{fig3}
\end{figure}
\end{center}

We first consider the acoustic transmission through the lenticular
structure of SCs. The results are presented in Fig.~\ref{fig3}.
The scenario is illustrated by Fig.~\ref{fig3}(a).  The two
dimensional spatial distribution of the normalized transmitted
intensity is shown in Fig.~\ref{fig3}(b). Here the $x$ axis is a
horizontal line towards right, and the $y$ axis is placed
vertically upward. We see that the focusing of the transmitted
wave is evident. Since the exact location of the maximum intensity
point is not clear from Fig.~\ref{fig3}(b), we have computed the
variation of the field intensity along the $x$ axis with $y=0$,
and also along the $y$ axis with $x=20$m. The results are shown in
Figs.~\ref{fig3}(c) and \ref{fig3}(d). Fig.~\ref{fig3}(c) shows
that there is a peak at $x \sim 20$m. The intensity variation
along the $y$ axis shows a significant peak at $y=0$. Here we
observe that the intensity is better confined along the $y$ axis
than along the $x$ axis. Some relevant parameters are listed in
the figure caption.

\input epsf.tex

Fig.~\ref{fig3} clearly demonstrate the acoustic converging
effects by SCs. Since the lenticular shape is thin, the focusing
effect may be quantified. According to the refraction theory, it
is possible to relate the object distance or length $s_o$, the
image distance $s_i$, and the focal length $f$ by the thin
formula\cite{Hecht}, \BE \frac{1}{s_o} + \frac{1}{s_i}
=\frac{1}{f}. \label{eq:thin} \EE If the medium can be regarded as
a refractive medium, the focal length will be related to the radii
of the curvatures of the convex surfaces by\cite{Hecht} \BE
\frac{1}{f} = (n-1)\left(\frac{1}{R_1} - \frac{1}{R_2}\right).
\label{eq:f}\EE Eqs.~(\ref{eq:thin}) and (\ref{eq:f}) lead to
Lensmaker's formula \BE \frac{1}{s_o} + \frac{1}{s_i} =
(n-1)\left(\frac{1}{R_1} - \frac{1}{R_2}\right),
\label{eq:lensmaker}\EE where $n$ is the refraction index.

We have verified that the acoustic imaging by the SCs discussed
above does obey Lensmaker's formula. First from the band
structure, we obtain the phase speed and subsequently the
refraction index as the ratio between the speeds in the air and in
the SCs. Then the focal length can be calculated from
Eq.~(\ref{eq:f}) with known $R_1$ and $R_2$. This will give the
theoretical value for the focal length. The focal length can also
be obtained by simulation. We vary the location of the source and
then find the image position. Taking these two values into
Eq.~(\ref{eq:thin}), we can get a simulated focal length for each
source location. If the device indeed behave as a lens, the
simulated focal lengths at various source locations should be
consistent. Furthermore, if the lens is refractive, the simulated
values should also be consistent with the theoretical value. We
found these claims are indeed valid for SCs.

\begin{center}
\begin{figure}[hbt]
\vspace{10pt} \epsfxsize=3in\epsffile{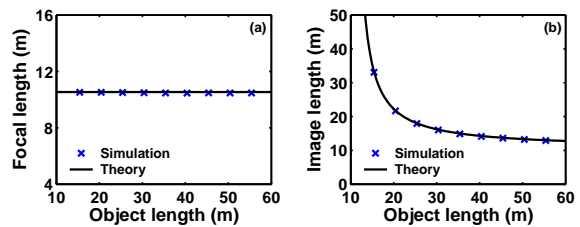} \caption{(a) The
focal lengths versus the object lengths; (b) image distance versus
the object length $s_o$. Both radii of the curvatures are taken as
4m. The theoretical results are shown by solid lines.}
\label{fig4}
\end{figure}
\end{center}

\begin{center}
\begin{figure}[hbt]
\vspace{10pt} \epsfxsize=2.5in\epsffile{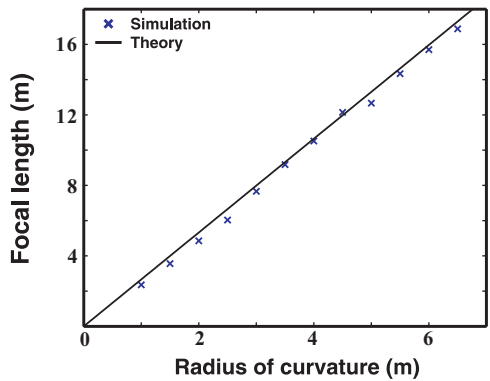} \caption{The
focal lengths versus the radius of curvature at frequency 1500Hz.
The crosses represent the simulation results, and the straight
line refers to the theoretical estimates.} \label{fig5}
\end{figure}
\end{center}

As an example, we consider the lenticular shape of arrays of
cylinders described in the figure caption of Fig.~\ref{fig3}. The
results are shown in Figure~\ref{fig4}. Here, both results from
the theoretical and simulations are presented. In (a), the focal
lengths estimated from the simulation for various object distances
are shown by the crosses. As the comparison, the theoretical focal
length calculated from Eq.~(\ref{eq:f}) is also shown. In (b), the
image distance versus the object distance is plotted. Here we see
that (1) from (a) the simulated focal lengths are consistent with
each other, and also consistent with the theoretical value; (2)
from (b) the relation between the image distance and the object
distance can be well described by Eq.~(\ref{eq:thin}) and also
agrees remarkably well with the theoretical curve.

We have further verified the above agreement by varying the radii
of the curvatures of the convex surfaces. The comparison with the
theoretical prediction is shown in Fig.~\ref{fig5}. In the
simulation, the thickness of the lens is kept unchanged at about
80 cm, when the radii are varied. We also take that both the radii
of the curved surfaces are the same; so the theoretical focal
length is $f = \frac{R}{2}\frac{1}{n-1}.$ The simulated value for
the focal length is obtained as follows. For a fixed radius of the
curvature, we vary the object distance and obtain the
corresponding image distance. Then the focal length can be
obtained through Eq.~(\ref{eq:thin}) for each object location. The
focal length will be averaged over all the locations. Then we
change to other values for the radius of the curvature. In this
way we obtain the simulated focal lengths as a functions of the
radius. The theoretical results are obtained from Eq.~(\ref{eq:f})
with the refraction index obtained from the band structures.
Figure~\ref{fig5} shows that the estimated focal lengths are in
excellent agreement with the theoretical prediction. Here we would
also like to note that the agreement will deteriorate when the
size of lenses is too small. In fact, when the size is too small,
the diffraction effect will become dominant and therefore the
focusing can no longer be described by the refraction process.

In summary, here we demonstrate that SCs not only can make
acoustic focusing devices, but also the focusing behavior can be
well described by Lensmaker's formula. This finding will help
design of actual acoustic refractive devices, such as focusing
audio speaker systems or sonars for large scale oceanographical
probing purposes \cite{Clay}. Underwater sonars with sonic crystal
lenses may improve the detection of fish \cite{Ye6} and other
marine organisms \cite{Chu}. Although the present work has been
focused upon the acoustic cases and on the rigid cylinders, it is
easy to see that the ideas can be readily applied to other SCs,
and as well as the photonic crystals.

\acknowledgments{This work received support from the National
Central University, National Science Council of Republic of
China.}

\end{document}